\documentclass[12pt]{iopart}

\usepackage[latin2]{inputenc}
\usepackage{setstack}
\usepackage{iopams}
\usepackage{amssymb}
\usepackage{epsfig}
\usepackage{bm}
\usepackage{graphicx}

\begin{document}

\title{Casimir-like force between intruders in granular gases}

\author{M.\ Reza Shaebani}

\address{Department of Theoretical Physics, Budapest University of 
Technology and Economics, H-1111 Budapest, Hungary}
\address{Institute for Advanced Studies in Basic Sciences, 
Zanjan 45195-1159, Iran}
\ead{shaebani@comphys.uni-duisburg.de}

\author{Jalal Sarabadani}

\address{Department of Physics, University of Isfahan, Isfahan 
81746, Iran}
\address{Institute for Studies in Theoretical Physics and
Mathematics, Tehran 19395-5531, Iran}
\ead{j.sarabadani@phys.ui.ac.ir}

\begin{abstract}
We numerically study a two-dimensional granular gas of rigid disks
where an external driving force is applied to each particle in
such a way that the system is driven into a steady state by
balancing the energy input and the dissipation due to inelastic
collisions. Two intruder particles embedded in this correlated
medium experience a fluctuation-induced force -- that is itself a
fluctuating quantity -- due to the confinement of the hydrodynamic
fluctuations between them. We find that the probability
distribution of this force is a Gaussian centered on a value that
is proportional to the steady-state temperature and grows
logarithmically with system size. We investigate the effect of the
other relevant parameters and estimate the force using the Fourier
transform of the fluctuating hydrodynamic fields.
\end{abstract}



\maketitle

\section{Introduction}
The \emph{Casimir force} predicted in the seminal work
of Casimir \cite{Casimir48} is an attractive force $ F= -\pi^2
\hbar c A / (240 D^4)$ between two perfect conducting neutral
plates with area $A$ facing each other at a distance $D$. This
attraction originates from the modification of the long-range
fluctuations of the quantum electromagnetic field, due to the
boundary conditions imposed by the conducting plates
\cite{Kardar99,Milonni94,Krech94,CasimirReview}. The Casimir 
energy -- the difference between the energies of the quantum
electromagnetic field for the plates at distance $D$ and at $D
\!\! \rightarrow \!\! \infty$, respectively -- is proportional to
$\hbar$ \cite{Casimir48,Kardar99,Milonni94}; so is the Casimir
force, that is the derivative of this energy with respect to the
distance between the plates.

Although the original Casimir interaction has a quantum nature,
such an effect occurs in many classical systems where fluctuations
are of thermal origin \cite{Kardar99}. In fact,
fluctuation-induced forces appear in systems with fluctuating
long-range correlations that are geometrically confined by
inserting external objects in the correlated medium. Examples can
be found in nematic liquid crystals \cite{Ziherl00}, critical
mixtures \cite{Fisher78,Hanke98}, superfluid films \cite{Ueno03},
and granular media \cite{Cattuto06,Brito07pre,Brito07gm}. In a 
thermally noisy correlated medium, where long range spatial 
correlations exist due to thermal fluctuations, the Casimir 
energy (and force) is expected to be proportional to $k_B T$ 
\cite{Kardar99}. The results of a recent experimental study 
on the critical mixtures reveal that the Casimir energy is 
indeed very sensitive to the temperature of the system 
\cite{Hertlein08}. Depending on the characteristics of 
the system, the fluctuation-induced force has been found to 
be even repulsive, e.g.\ in dielectric materials with nontrivial 
magnetic susceptibility \cite{Kenneth02}, or the interaction 
between a perfectly conducting and an infinitely permeable plate \cite{Boyer74}.

The fluctuation-induced force between two intruder objects
immersed in a granular gas is studied in \cite{Cattuto06}, where
it is found that the confinement of the fluctuation spectrum of
the hydrodynamic pressure field induced by the intruders, leads to different
local pressures in the gap between the intruders and the outside
region. This effect causes an effective Casimir-like force between
the intruders. The results of some experiments on granular fluids
have confirmed the idea that the presence of large intruder
particles modifies the thermodynamic properties such as pressure,
velocity and density in the regions between the larger particles
\cite{Sanders04,Aumaitre01,Zuriguel05}. Appearance of long-range
interactions (despite the short-range nature of the interactions
on the grain scale) may shed new light on the mechanism of some
collective behaviors, e.g.\ segregation
\cite{Aumaitre01,Zuriguel05}, in granular media.

In this paper we present the results of extensive numerical
simulations to study the fluctuation-induced force between two
large objects immersed in a noisy granular gas. Although the
interactions between particles in granular systems are dissipative
\cite{Jaeger96}, we maintain the dynamics by means of an external
driving force. Our main aims in the present work are to
investigate the probability distribution of the fluctuating
Casimir-like force and the relationship between the average force
and important parameters of the system such as the steady-state
temperature, the mass density of gas particles and the distance
between the intruders. The question of whether the average force
depends on the system size is also addressed.

\section{Simulation Method}
\begin{figure}
\centering
\includegraphics[scale=0.5,angle=0]{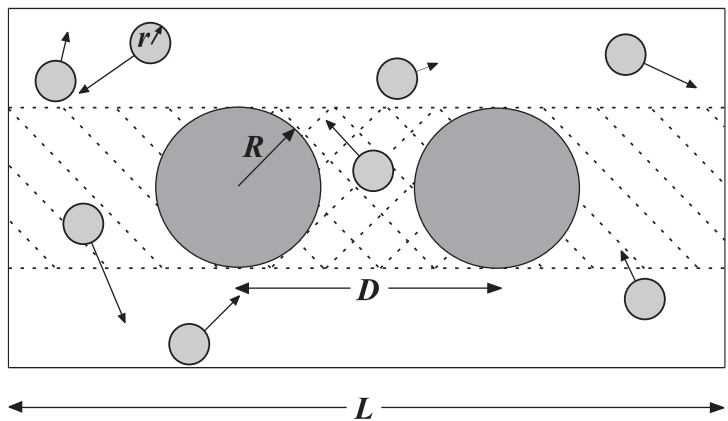} 
\caption{Schematic picture showing the simulation cell. The pressure 
difference between the hatched and cross-hatched regions leads to an
effective repulsive force between the two intruders. We choose
$L_{_0}/r_{_0}\!\! =\! 200$, $D_{_0}/r_{_0}\!\! =\! 30$ and
$R_{_0}/r_{_0}\!\! =\! 10$ in the reference system.} 
\label{Fig-Schematic}
\end{figure}

The system we consider is a two-dimensional granular gas of 
identical rigid disks of radius $r$ and mass $m$ interacting via
inelastic collisions. In order to exclude the undesired effects 
of side walls, periodic boundary conditions are applied in 
both directions of the square-shaped system of length $L$. Two
immobile rigid intruder particles of radius $R$ and infinite mass
and moment of inertia are immersed in the granular gas bed, 
separated by a distance $D$. We have one reference 
system whose parameters are denoted by zero subscripts (see
Fig.~\ref{Fig-Schematic}). Throughout the paper we either use this
reference system or vary only one parameter to check its effect
while other parameters are kept fixed at their reference values.
We perform molecular dynamics simulations in which the number 
density $n$ of the granular gas is $0.075r^{-2}$ and the effective
normal coefficient of restitution $\alpha$ of the particle-particle 
and particle-intruder collisions is set to 0.8.

The system is coupled to an external heat bath 
that homogeneously transfers energy into the system. The
acceleration of each particle $\bm a_i$ is perturbed
instantaneously according to $\bm a_i^\prime  = \bm a_i + \bm
\xi_i$, where prime refers to the acceleration after perturbation, 
and $\bm \xi_i$ can be considered as Gaussian 
white noise with zero mean and correlation $\langle \xi_{i \alpha}(t) 
\xi_{j \beta}(t^{\prime})\rangle = \xi_{_0}^2 \delta_{ij} 
\delta_{\alpha \beta} \delta(t-t^{\prime})$, where $\alpha$ 
and $\beta$ denote Cartesian components of vectors 
\cite{vanNoije99,Peng98}. Practically, the energy is
transferred into the system in the following way
\cite{Cattuto06,vanNoije99}: we update the momentum of each grain
at each time step $\Delta t$. The components of the momentum are
updated by adding random values that are chosen from a Gaussian
distribution with zero mean and variance $\sigma^2_{_0}$.

\section{Results}

\begin{figure}
\centering
\includegraphics[scale=1.0,angle=0]{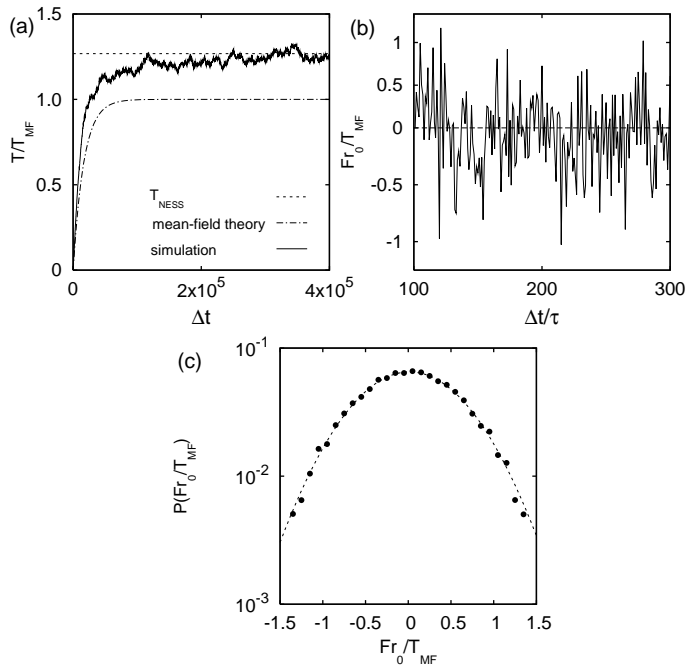} \caption{(a)
Temperature saturation starting from a random initial
configuration of the system (solid line). The temperature $T$ is
scaled by the mean-field steady-state temperature
$T_{_{MF}}$. The dash-dot line shows the time evolution of
$T$ according to Eq.~(\ref{TemperatureParameters}). The
nonequilibrium stationary temperature $T_{_{NESS}}$ is
shown with dashed line. (b) The time evolution of
fluctuation-induced force $F$ scaled by $T_{_{MF}} /
r\!_{_0}$. (c) The probability distribution of $F$ for the
reference system where the simulation run over $10^8$ time steps.
The dashed line shows the best fit with a Gaussian distribution
centered on $0.015 T_{_{MF}} \/ r_{_0}$.}
\label{Fig-ForceDistribution}
\end{figure}

The simulation results [solid line in 
Fig.~\ref{Fig-ForceDistribution}(a)] reveal that the temperature 
$T$ is time dependent and finally saturates. Note that the 
temperature $T$ is a uniform field in our simulations in 
contrast to the case where energy flows into the system 
from the boundaries which leads to the spatial gradient of
temperature \cite{Du95}. We follow a mean-field approach to 
describe the time evolution of $T$. On the one hand the 
system gains energy due to coupling with the heat bath. 
The rate of the energy gain of a single particle averaged 
over the uncorrelated noise source $g(\xi)$ is
\begin{equation}
   \partial_t E \! = \! \lim_{\Delta t \rightarrow 0} \frac{1}{\Delta t}
   \int_{\xi}  [ E_i(t+\Delta t)-E_i(t) ] g(\xi) d\xi = dm\xi_{_0}^2.
\end{equation}  
On the other hand the system losses energy due to inelastic 
collisions. The rate of the energy loss of a single particle 
is described by $\partial_t E = -(1-\alpha^2)\omega T /d = \beta
T^{3/2}$ \cite{vanNoije97,vanNoije98}, where $d$ is the dimension
of the system, $\omega (\propto \!\sqrt T)$ is the
temperature-dependent collision frequency given by the Enskog
theory \cite{Chapman70}, and $\beta$ is a coefficient that
contains all of the relevant parameters except the temperature. 
Therefore the time dependence of temperature (energy) 
according to mean-field approximation is given by 
$dT/dt = -\beta T^{3/2} + dm\xi_{_0}^2$. By integrating the equation 
$dT/(-\beta T^{3/2} + dm\xi_{_0}^2) = dt$ \cite{Gradshteyn}, one arrives at the 
following expression for the evolution of $T$ [dash-dot line in 
Fig.~\ref{Fig-ForceDistribution}(a)]:
\begin{equation}
f(T) - f(T_i) = -dm \xi_{_0}^2 t/2T_{_{MF}},
\end{equation}
where $T_i$ is the initial temperature and
\begin{equation}
f(x) = \frac{1}{6} \ln {\frac{1 - 2 {\sqrt{x /T_{_{MF}}}} + x/T_{_{MF}}} 
{1 + {\sqrt{x/T_{_{MF}}}} + x/T_{_{MF}}}} + 
{\frac{1}{\sqrt 3}} \arctan 
{\frac{2 {\sqrt{x/T_{_{MF}}}} + 1}{\sqrt 3}}.
\end{equation}

Starting from any initial configuration, the driven granular gas
finally converges to a nonequilibrium steady state, where energy
dissipation due to inelastic collisions is balanced with the
energy input. Although the mean-field prediction for the saturated
temperature is $T_{_{MF}} \!\!=\!\! 
(dm\xi_{_0}^2/\beta)^{2/3}$, the temperature of the nonequilibrium
steady state $T_{_{NESS}}$ [dashed line in 
Fig.~\ref{Fig-ForceDistribution}(a)] is expected to be larger than 
$T_{_{MF}}$ since it is logarithmically divergent in
the system size (in 2D) due to the existence of spatial 
hydrodynamic fluctuations \cite{vanNoije99}. 

After the system achieves the stationary state, we study the
effective interaction between the two fixed intruders. We measure
the total momenta $\bm P_\ell$ and $\bm P_r$ transferred from the
granular gas to the left and right intruders respectively, during
a time interval $\tau$. $\tau$ corresponds to $10^4$ time steps
($\sim 100$ collisions per grain). The components of $\bm P_\ell$
and $\bm P_r$ parallel to the line connecting the two centers of
intruders provide the fluctuation-induced force as $F\!\!=\!\!
(P_{x r}\! -\!\! P_{x \ell} ) / 2 \tau$. Since this force is
itself a fluctuating quantity, as expected for such correlated
media \cite{Bartolo02}, we let the simulation run until we measure
the force for more than $10^4$ consecutive time intervals $\tau$.
Figure \ref{Fig-ForceDistribution}(b) displays the fluctuating
nature of $F$ when measured in the steady state of the reference
system. The probability distribution of $F$ is plotted in
Fig.~\ref{Fig-ForceDistribution}(c), where it turns out that the
distribution can be well fitted to a Gaussian with the standard
deviation $0.612 T\!_{_{MF}}/r\!_{_0}$. The ensemble average
of $F$ in the reference system equals to $\langle F \rangle\!_{_0}
= 0.015 T\!_{_{MF}}/r\!_{_0}$. The fluctuations are about
two orders of magnitude larger than $\langle F \rangle\!_{_0}$, 
thus very long simulations are required to measure the force with 
small numerical errors. Such a repulsive force between two intruder 
objects sitting at a distance larger than their radii was reported 
in the previous studies of driven granular gases 
\cite{Cattuto06,Brito07gm} and even dense shaken granular 
packings \cite{Duran98}.

\begin{figure}
\centering
\includegraphics[scale=1.3,angle=0]{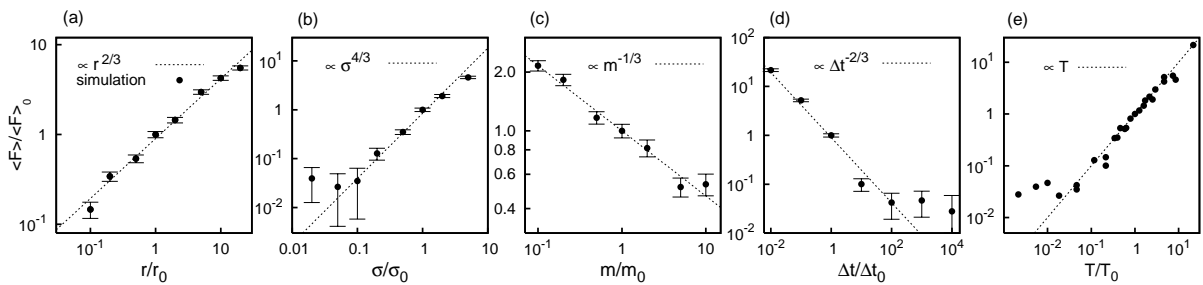} \caption{The
fluctuation-induced force $\langle F \rangle$ scaled by $\langle F
\rangle_{_0}$ in terms of the grain radius $r\!$ (a), the standard
deviation of the momentum distribution $\sigma$ (b), the grain
mass $m$ (c), and the time step $\Delta t$ (d). The dashed lines
are power-law functions with exponents correspond to those of
Eq.~(\ref{TemperatureParameters}). (e) $\langle F \rangle$ versus
the temperature $T$. Here, the dashed line displays a linear
growth of $\langle F \rangle$ with $T$.}
\label{Fig-ForceParameters}
\end{figure}

The time step length -- that reflects the time scale for 
interaction with the heat bath -- is chosen large enough that 
the dissipation is kept alive, but it is much smaller than the mean 
free time between the collisions. Since the energy considerations 
yield $\xi_{_0}^2 = \sigma_{_0}^2/(m^2 \Delta t)$ \cite{vanNoije99}, 
the steady state temperature (according to the Enskog theory) then 
becomes  
\begin{equation}
 T_{_{NESS}} \propto \bigg( \frac{ r ~ \sigma^2}{
 \sqrt{m} ~ \Delta t } \bigg)^{2/3}. \label{TemperatureParameters}
\end{equation}
To investigate how the Casimir-like force is affected by the
simulation parameters, we vary the parameter values one by one
while the others are kept fixed at their reference values and
measure the force. For this purpose, simulations are performed
anew for each data point in Figs.~\ref{Fig-ForceParameters}(a-d)
and the force is measured after the system reaches the steady
state. We note that the values of $\langle F \rangle\!/\!\langle F
\rangle\!_{_0}$ around $5\! \times\!\! 10^{-2}$ reflect the
accuracy level of our calculations. Therefore, the best fits with
power-law functions (not shown in Fig.~\ref{Fig-ForceParameters})
are obtained when the inaccurate data in
Figs.~\ref{Fig-ForceParameters}(b) and (d) are neglected. Our
results reveal that the exponents of the best fits approximately
equal to the exponents of Eq.~\ref{TemperatureParameters} with
less than $\% 6$ errors. The stationary temperature is also
measured for each data point in
Figs.~\ref{Fig-ForceParameters}(a-d). The average force is plotted
as a function of the temperature in
Fig.~\ref{Fig-ForceParameters}(e) that interestingly verifies the
linear dependence of the fluctuation-induced force on the
temperature, as theoretically was predicted in the literature for
the thermal fluctuating correlated media
\cite{Kardar99,Krech94}.

\begin{figure}
\centering
\includegraphics[scale=1.0,angle=0]{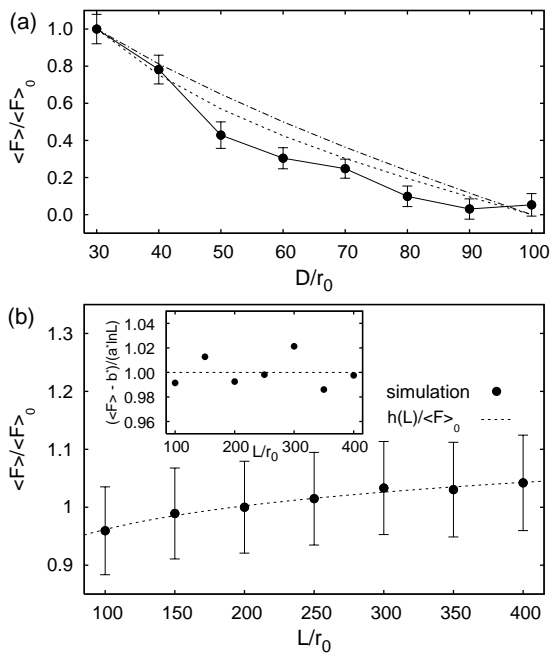} \caption{(a) The average
force $\!\langle F \rangle\!$ scaled by $\!\langle F
\rangle\!_{_0}\!$ in terms of the distance between the intruders
$D$ scaled by $\!r\!_{_0}$. The dashed (dash-dot) line corresponds
to the force calculation with (without) taking into account the
actual boundary conditions. (b) $\!\langle F \rangle\! / 
\!\langle F \rangle\!_{_0}\!$ as a function of the system size $L$. 
The dashed line corresponds to ${h}(L)$ scaled by 
$\!\langle F \rangle\!_{_0}\!$. The inset shows how the deviation of 
the data from the logarithmic growth ${h}(L)$ varies with the system size.} 
\label{Fig-ForceDistanceSystemSize}
\end{figure}

Next we investigate how the effective interaction between the
intruders is affected by the distance between them. Figure 
\ref{Fig-ForceDistanceSystemSize}(a) displays that the average
repulsive force $\langle F \rangle$ decays with increasing the
distance $D$. Supposing a power-law dependence of $\langle F
\rangle$ on $D$, one gets the exponent $-0.8 \pm 0.3$.

The repulsive force originates from the pressure difference 
between the hatched and cross-hatched regions in Fig.~\ref{Fig-Schematic}. 
The mechanism is briefly explained below (for details see 
\cite{Cattuto06}). In a system of hard disks, the 
pressure field $p$ can be written as $p(n,T) = T H(n)$ \cite{Verlet82}, 
where $n$ and $T$ are fluctuating hydrodynamic fields. 
Expanding the pressure up to the second order 
around the stationary values ($n_0,T_0$) and taking its
statistical average over the random noise source, we obtain 
$\langle p \rangle = p_{_0} \!+\! H_1 \langle \delta n 
\delta T \rangle \!+\! T_{_{MF}} H_2 \langle \delta n^2 \rangle$, 
where $p_{_0}$ is the stationary pressure and $H_1\!$ and $\!H_2\!$ are the
first and second derivatives of $H\!(n)$ with respect to $n$
around $n_0$, respectively. By employing the Fourier 
transform of the fluctuating fields $\delta A(\bm r) \!=\!\! 
\sum_k e^{-i\bm k \cdot \bm r} \delta A_{\bm k}/\!V$ and the 
structure factors $S_{AB}(\bm k) \! = \!\langle \delta A_{\bm k}
\delta B_{\!-{\bm k}} \!\rangle\!/\!V$ \cite{vanNoije99} we 
rewrite $p$ \cite{Cattuto06}:
\begin{equation}
\langle p \rangle = p_{_0} \!+\! V^{-1} \!\! \sum_{\bm k}
[H_1 S_{nT}\!(\bm k) \! + \! T_{_{MF}} H_2 S_{nn}
\!(\bm k) ]. \label{PressureStructurFactor}
\end{equation}
In the calculation of the pressure the main contribution comes 
from the region of small $\bm k$. Substituting the structure 
factors of this region, $S_{AB}(\bm k)=S_{AB}^0 /k^2$ 
(where $S_{AB}^0$ is a function of the number density, 
restitution coefficient and noise intensity), we obtain the 
steady-state pressure: $<p>= p_{_0} + CV^{-1} \sum_k (1/k^{2})$, 
where $C = H_1 S_{nT}^0 + T_{_{MF}} H_2 S_{nn}^0$ 
is a function of the number density $n$ and is negative for our 
simulations. In the gap between the intruders, the number of 
valid $k$ modes (and therefore $\sum_k 1/k^2$) is smaller 
than the outside region. Consequently, the pressure in the 
cross-hatched region is higher than the hatched region. This 
pressure difference $\Delta p$ causes an effective repulsive 
force between the intruders. Interestingly, $C$ (and therefore 
$\Delta p$ and the fluctuation-induced force) is proportional 
to $T_{_{MF}}$.

$\Delta p$ is estimated in Ref.~\cite{Cattuto06}, supposing that
$k$ vectors are confined in rectangular boxes of size
$(D\!-\!2R)\!\times\! 2R$ and $(L\!-\!D\!-\!2R)\!\times\! 2R$ in
the gap between the intruders and in the outside region,
respectively. The dash-dot line in 
Fig.~\ref{Fig-ForceDistanceSystemSize}(a) displays the Casimir 
force for different distances between the intruders, calculated
according to the above mentioned estimation. The force is 
overestimated because (i) the geometrical simplification (using 
rectangular instead of circular boundary conditions) causes an 
error in the computed pressure, and (ii) the fluctuations are 
indeed correlated in the hatched and cross-hatched regions. We have 
improved this estimation by taking into account the circular shape 
of the intruders. At each point in the hatched or cross-hatched 
regions, the components of the allowed $\!\bm k\!$ vectors in the 
$x$ and $y$ (perpendicular to $x$) directions are $( 2 \pi 
n_x/d_x\!(y) , 2 \pi n_y/L )$, where $d_{x} \!(y)\!$ is the 
length of the line segment that horizontally connects the surfaces 
of the two intruders. Using the continuous form of $\sum_k 1/k^2$, 
the pressure at this point reads
\begin{equation}
\langle p \rangle  = p_{_0} \!+\! CV^{-1} \frac{(2 \pi)^2}{L d_x} \!
\! \int_{2\pi/d_x}^{2\pi/r_{_0}} \! \! dk_x \! \!
\int_{2\pi/L}^{2\pi/r_{_0}} \! \! \! dk_y   \frac{1}{k_x^2+k_y^2},
\label{PressureNew}
\end{equation}
Where an upper cutoff $2 \pi / r_{_0}$ is used beyond which 
hydrodynamics is not valid. This cutoff is also needed to exclude 
the singular vector $\bm k \!=\! (0,0)$ from the integral. Calculating the pressure 
in the hatched and cross-hatched regions at a given $y$ [according 
to Eq.~(\ref{PressureNew})], we obtain $\Delta p(y)$. Next, 
by integrating $\Delta p(y)$ over the whole range of
$y$ ($-R,R$) to obtain the average force, and repeating the process 
for different distances between the intruders we get the dashed 
line in Fig.~\ref{Fig-ForceDistanceSystemSize}(a), revealing a 
significant improvement due to taking into account the actual 
boundary conditions. We note that the correlation between the 
fluctuations in the two regions seems to be important since our 
results still overestimate the force. One expects that the 
interaction between intruders would be affected if a new 
intruder is inserted into the system. Therefore, our results 
provides an important evidence for the nonlinearity of the 
Casimir-like forces in thermally noisy systems and offers 
new insight into the problem which should trigger new 
experiments and theoretical explanations.

Finally we study the system-size dependence of
the effective force. It has been found in
Ref.~\cite{vanNoije99} that $T_{NESS} - T_{MF}$ in a 
two-dimensional driven granular gas is logarithmically
divergent in the system size due to hydrodynamic 
fluctuations. Such a size dependence is also expected 
for $\langle F \rangle$ since we have verified a linear 
dependence of $\langle F \rangle$ on $T_{NESS}$. 
The simulation results [Fig.~\ref{Fig-ForceDistanceSystemSize}(b)] 
reveal that $\langle F \rangle$ increases slightly as the system size $L$ 
increases and the data can be well fitted to a logarithmic curve 
${h}(L) = {a}^* {ln} L + {b}^*$ with less than $\% 4$ error. Within the 
investigated range of $L$, the deviation of the data from the logarithmic 
relation does not show a systematic dependence on the system size [inset of 
Fig.~\ref{Fig-ForceDistanceSystemSize}(b)]. The increase of $\langle F \rangle$ with $L$ can 
be also explained qualitatively in the following way: When
the system size is increased at a fixed $D$, the number of
possible $k$ modes (and $\sum_k 1/k^2$) in the hatched region of
Fig.~\ref{Fig-Schematic} increases; consequently, the mean pressure
in the hatched region is reduced while the pressure in the cross-hatched 
region is not varied. This leads to the increase of pressure 
difference between the hatched and cross-hatched regions which
yields the increase of $\langle F \rangle$.

\section{Conclusions}
In summary, we have carried out numerical simulations of the
fluctuation-induced force between two intruders in a driven
granular gas bed. Earlier works have demonstrated that such a
force is expected since the thermodynamic properties are affected
due to the presence of the intruders. Here, our main focus has
been to study the fluctuating nature of this interaction and the
temperature dependence of the ensemble average of the force
$\langle F \rangle$. We have verified that $\langle F \rangle$
increases linearly with temperature, and slightly with system 
size that can be described by a logarithmic growth. We
have also improved the estimation given in \cite{Cattuto06} to
explain the force, by taking into account the actual boundary
conditions of the problem.

\section{Acknowledgments}
We would like to express our gratitude to T. Unger for valuable 
discussions and B. Farnudi for reading the manuscript. J.S.
acknowledges the Department of Physics of the Institute for Advanced
Studies in Basic Sciences for the hospitality.


\section*{References}
\begin{thebibliography}{10}
\bibitem{Casimir48} Casimir H.~B.~G., 1948 {\it Proc. K. Ned. Akad. Wet.} \textbf{51} 793
\bibitem{Kardar99} Kardar M. and Golestanian R., 1999 {\it Rev. Mod. Phys.} \textbf{71} 1233
\bibitem{Milonni94} Milonni P., 1994 {\it The Quantum Vacuum} (San Diego: Academic)
\bibitem{Krech94} Krech M., 1994 {\it The Casimir Effect in Critical Systems} (Singapore: World Scientific)
\bibitem{CasimirReview} Plunien G., M\"uller B. and Greiner W., 1986 {\it Phys. Rep.} \textbf{134} 87

\hspace{-0.35cm}Bordag M., Mohideen U. and Mostepanenko V. M., 2001 {\it Phys. Rep.} \textbf{353} 1

\hspace{-0.35cm}Milton K. A., 2004 {\it J. Phys. A: Math. Gen.} \textbf{37} R209

\hspace{-0.35cm}Lamoreaux S. K., 2005 {\it Rep. Prog. Phys.} \textbf{68} 201

\hspace{-0.35cm}Mostepanenko V. M. and Turnov N. N., 1997 {\it The Casimir Effect and its Applications} (Oxford: Clarendon)

\hspace{-0.35cm}Milton K. A., 2001 {\it The Casimir Effect: Physical Manifestations of the Zero-Point Energy} (Singapore: World Scientific)
\bibitem{Ziherl00} Ziherl P., Podgornik R. and Zumer S., 2000 {\it Phys. Rev. Lett.} \textbf{84} 1228
\bibitem{Fisher78} Fisher M. E. and de~Gennes P. G., 1978 {\it C. R. Acad. Sci. Paris B} \textbf{287} 207
\bibitem{Hanke98} Hanke A. et al., 1998 {\it Phys. Rev. Lett.} \textbf{81} 1885
\bibitem{Ueno03} Ueno T. et al., 2003 {\it Phys. Rev. Lett.} \textbf{90} 116102
\bibitem{Cattuto06} Cattuto C. et al., 2006 {\it Phys. Rev. Lett.} \textbf{96} 178001
\bibitem{Brito07pre} Brito R., Marconi U. M. B. and Soto R., 2007 {\it Phys. Rev. E} \textbf{76} 011113
\bibitem{Brito07gm} Brito R., Soto R. and Marconi U. M. B, 2007 {\it Granular Matter} \textbf{10} 29
\bibitem{Hertlein08} Hertlein C. et al., 2008 {\it Nature} \textbf{451} 172
\bibitem{Kenneth02} Kenneth O. et al., 2002 {\it Phys. Rev. Lett.} \textbf{89} 033001
\bibitem{Boyer74} Boyer T. H., 1974 {\it Phys. Rev. A} \textbf{9} 2078 
\bibitem{Sanders04} Sanders D. A. et al., 2004 {\it Phys. Rev. Lett.} \textbf{93} 208002
\bibitem{Aumaitre01} Aumaitre S., Kruelle C. A. and Rehberg I., 2001 {\it Phys. Rev. E} \textbf{64} 041305
\bibitem{Zuriguel05} Zuriguel I. et al., 2005 {\it Phys. Rev. Lett.} \textbf{95} 258002
\bibitem{Jaeger96} Jaeger H. M., Nagel S. R. and Behringer R. P., 1996 {\it Rev. Mod. Phys.} \textbf{68} 1259
\bibitem{vanNoije99} van Noije T. P. C. et al., 1999 {\it Phys. Rev. E} \textbf{59} 4326
\bibitem{Peng98} Peng G. and Ohta T., 1998 {\it Phys. Rev. E} \textbf{58} 4737
\bibitem{Du95} Du Y., Li H. and Kadanoff L. P., 1995 {\it Phys. Rev. Lett.} \textbf{74} 1268
\bibitem{vanNoije97} van Noije T. P. C. et al., 1997 {\it Phys. Rev. Lett.} \textbf{79} 411
\bibitem{vanNoije98} van Noije T. P. C., Ernst M. H. and Brito R., 1998 {\it Phys. Rev. E} \textbf{57} R4891
\bibitem{Chapman70} Chapman S. and Cowling T. G., 1970 {\it The Mathematical Theory of Non-uniform Gases} (Cambridge: Cambridge University Press)
\bibitem{Gradshteyn} Gradshteyn I. S. and Ryzhik I. M., 2007 {\it Table of Intergals, Series and Products} (Amsterdam: Academic Press) p 72
\bibitem{Bartolo02} Bartolo D. et al., 2002 {\it Phys. Rev. Lett.} \textbf{89} 230601
\bibitem{Duran98} Duran J. and Jullien R., 1998 {\it Phys. Rev. Lett.} \textbf{80} 3547
\bibitem{Verlet82} Verlet L. and Levesque D., 1982 {\it Mol. Phys.} \textbf{46} 969
\end{thebibliography}
\end{document}